\documentclass[14pt]{extarticle}
\usepackage[a4paper,total={18cm,27cm}]{geometry}

\usepackage[utf8]{inputenc}
\usepackage{amsmath}
\usepackage{amsfonts}

\usepackage{graphicx}
\usepackage{array}
\usepackage{caption}

\usepackage{xcolor}
\colorlet{shadecolor}{yellow!20}

\usepackage{authblk}

\usepackage{hyperref}

\newcommand{\NI}{\vspace{0.2cm}\noindent}

\begin{document}

% **************************************************
% TITLE
% **************************************************

\title{Organizational Regularities\\ in Recurrent Neural Networks}

% **************************************************
% AUTORS
% **************************************************

\author[1]{Claus Metzner}
\author[1]{Achim Schilling}
\author[1]{Andreas Maier}
\author[1]{Patrick Krauss}

\affil[1]{\small Cognitive Computational Neuroscience Group, Pattern Recognition Lab, Friedrich-Alexander-University Erlangen-Nürnberg (FAU), Germany}

\maketitle

% ***************************************************
% ABSTRACT
% ***************************************************

\begin{abstract}
\NI Previous work has shown that the dynamical regime of Recurrent Neural Networks (RNNs)—ranging from oscillatory to chaotic and fixpoint behavior—can be controlled by the global distribution of weights in connection matrices with statistically independent elements. However, it remains unclear how network dynamics respond to organizational regularities in the weight matrix, as often observed in biological neural networks. Here, we investigate three such regularities: (1) monopolar output weights per neuron, in accordance with Dale's principle, (2) reciprocal symmetry between neuron pairs, as in Hopfield networks, and (3) modular structure, where strongly connected blocks are embedded in a background of weaker connectivity. We construct weight matrices in which the strength of each regularity can be continuously tuned via control parameters, and analyze how key dynamical signatures of the RNN evolve as a function of these parameters. Moreover, using the RNN for actual information processing in a reservoir computing framework, we study how each regularity affects performance. We find that Dale monopolarity and modularity significantly enhance task accuracy, while Hopfield reciprocity tends to reduce it by promoting early saturation, limiting reservoir flexibility.
\end{abstract}

% *******************************************************
\newpage

\section{Introduction}

\NI Over the past decades, deep learning has achieved remarkable progress \cite{lecun2015deep,alzubaidi2021review}, notably through the rise of large language models \cite{min2023recent}. These models are typically based on feedforward architectures, where information flows unidirectionally from input to output layers. In contrast, Recurrent Neural Networks (RNNs) include feedback connections, enabling them to function as autonomous dynamical systems \cite{maheswaranathan2019universality} that sustain neural activity even without ongoing external input.

\NI RNNs exhibit certain "universal" properties—such as the ability to approximate arbitrary functions \cite{schafer2006recurrent} or general dynamical systems \cite{aguiar2023universal}—which, alongside other strengths, have spurred interest in their fine-grained behavior. For example, they can preserve information from temporally extended input sequences \cite{jaeger2001echo,schuecker2018optimal,busing2010connectivity,dambre2012information,wallace2013randomly,gonon2021fading} and learn effective internal representations by balancing compression and expansion of information \cite{farrell2022gradient}.

\NI A further key research theme concerns the control of RNN dynamics, including how internal and external noise shape network behavior \cite{rajan2010stimulus,jaeger2014controlling,haviv2019understanding,molgedey1992suppressing,ikemoto2018noise,krauss2019recurrence,bonsel2021control,metzner2022dynamics}. RNNs have also been proposed as models for neural computation in the brain \cite{barak2017recurrent}. Notably, sparse RNNs—with a low average node degree, resembling biological circuits \cite{song2005highly}—have shown improved capacity for information storage \cite{brunel2016cortical,narang2017exploring,gerum2020sparsity,folli2018effect}.

\NI In our earlier work, we systematically explored the interplay between network structure and dynamics, beginning with three-neuron motifs \cite{krauss2019analysis}. We later showed how the weight distribution’s width $w$, the connection density $d$, and the balance $b$ between excitation and inhibition can be tuned to control the dynamics of large, autonomously active RNNs \cite{krauss2019weight,metzner2022dynamics}. We also investigated noise-induced resonance phenomena in such systems \cite{bonsel2021control,schilling2022intrinsic,krauss2016stochastic,krauss2019recurrence,schilling2021stochastic,schilling2023predictive,metzner2024recurrence}.

\NI Most studies—including our own—have assumed statistically independent weight matrix elements, drawn from fixed distributions and assigned randomly. While analytically convenient, this assumption does not reflect the structural regularities seen in biological networks. Real neural systems exhibit highly non-random connectivity, shaped by development, functional demands, and evolutionary constraints. 

\NI The potential benefits of incorporating structural regularities into neural architectures have increasingly been recognized in the literature. Several recent developments have explored this direction, aiming to move beyond fully random or uniform connectivity. For example, Capsule Networks introduced by Hinton and colleagues \cite{sabour2017dynamic,hinton2018matrix} implement local groups of neurons—capsules—that preserve part-whole relationships via structured routing mechanisms. In transformer models, architectural variants have been proposed that introduce explicit modularity or routing constraints to enhance interpretability and scalability \cite{rosenbaum2017routing,shazeer2017outrageously,rosenbaum2019routing}. Moreover, a number of studies have examined recurrent networks with biologically inspired topology, exploring the impact of modularity, reciprocity, or Dale-like constraints on network dynamics and learning performance \cite{zador2019critique,kim2021neural,chang2022modularity}.

\NI These efforts highlight a growing consensus that structural features—long regarded as biological idiosyncrasies—may in fact play a functional role in shaping the computational behavior of artificial neural systems. Against this background, we systematically examine the isolated effects of three such regularities:

\NI First, biological neurons follow Dale’s Principle, meaning each neuron is either excitatory or inhibitory, but not both \cite{strata1999dale,somogyi1998salient}.

\NI Second, neural circuits exhibit an increased likelihood of reciprocal connections: if neuron A projects to B, B is more likely to project back to A \cite{song2005highly,perin2011synaptic}. This bidirectional coupling introduces local symmetries that may stabilize attractor states and support mutual reinforcement.

\NI Third, the brain is modular, comprising groups of neurons more densely connected within than between groups \cite{sporns2016modular,meunier2010modular}. Such organization appears across scales, from cortical microcircuits to large-scale areas, and enables specialized yet integrative processing.

\begin{figure}[ht!]
%\centering
\includegraphics[width=0.95\linewidth]{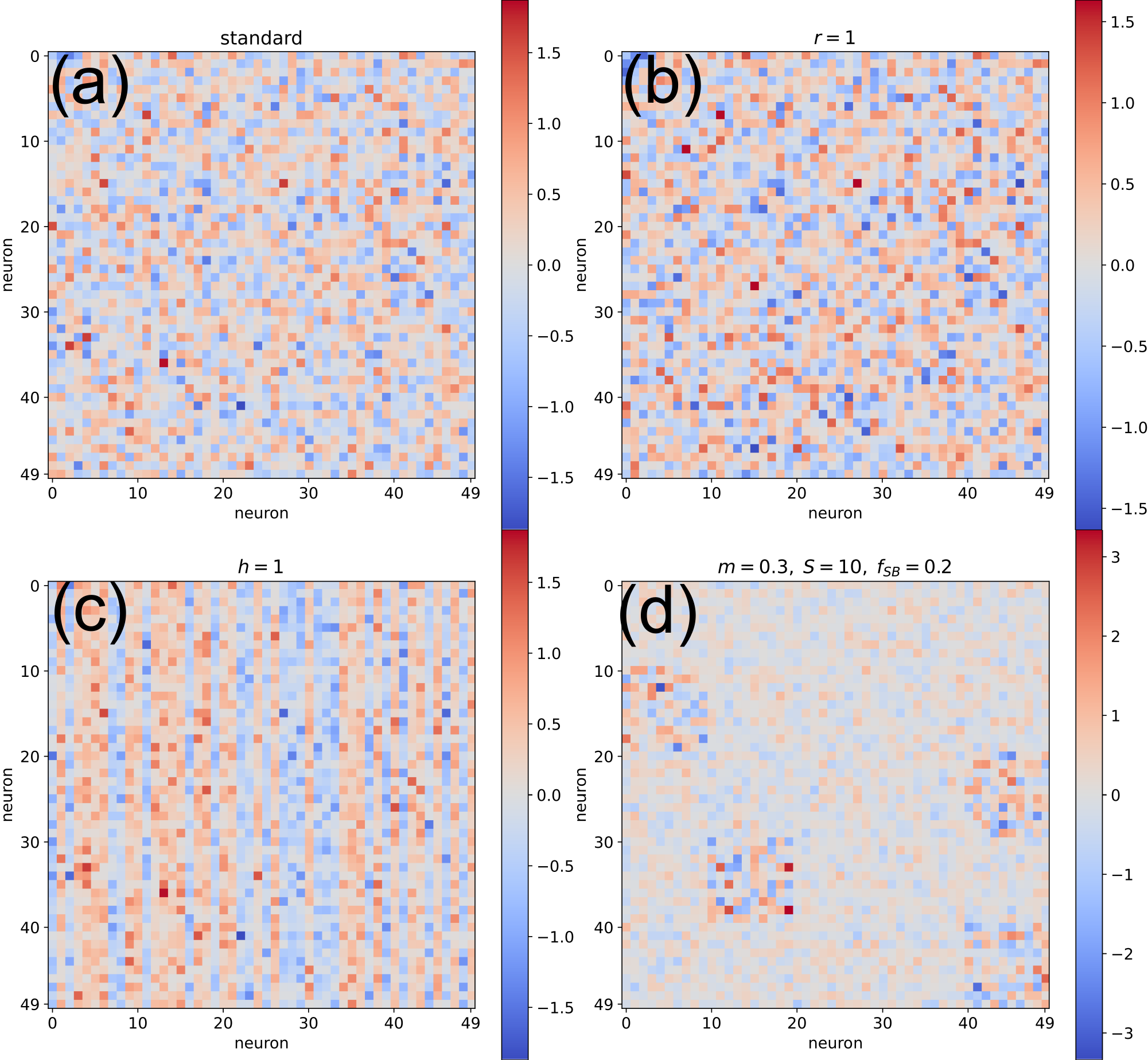}
\caption{{\bf Example weight matrices}. \textbullet$\;${\bf(a)} A standard weight matrix of size $50\times50$, with density $d=1$, balance $b=0$, width $w=0.5$, and all organizational regularity parameters set to zero: $r=h=m=0$.
\textbullet$\;${\bf(b)} With maximal Hopfield reciprocity $r=1$ and all other parameters unchanged, the matrix becomes symmetric about the diagonal, while preserving density, balance, and width.
\textbullet$\;${\bf(c)} With maximal Dale homogeneity $h=1$, each matrix column (representing a neuron's output weights) adopts a uniform sign.
\textbullet$\;${\bf(d)} With modularity $m=0.3$, block size $S=10$, and probability $f_{SB}=0.2$, some blocks exhibit significantly increased fluctuation width, while others become weaker. Note the different color bar in (d).}
\label{fig_MatPlt}
\end{figure}

\NI In the following, we study how each of these biologically inspired regularities affects the dynamical regime of RNNs, using a minimal implementation in which their respective strength can be continuously tuned (compare Fig.\ref{fig_MatPlt}).

\NI The first regularity, called {\bf Dale Homogeneity}, is controlled via a continuous parameter $h\in[0,1]$, indicating the degree to which a neuron's outputs are consistently excitatory or inhibitory. At $h=0$, polarities are mixed; at $h=1$, all neurons are strictly monopolar.

\NI The second regularity, {\bf Hopfield Reciprocity}, is controlled by a parameter $r\in[0,1]$. For $r=0$, weights $w_{ij}$ and $w_{ji}$ are independent; for $r=1$, they are identical, as in Hopfield networks.

\NI The third regularity is {\bf Modularity}, parameterized by $m\in[0,1]$. A network with $m=0$ lacks modular structure, whereas $m>0$ introduces strongly connected quadratic blocks with standard deviation $w_S > w$, embedded in a weaker background with $w_W < w$. As the degree of modularity $m$ increases, $w_S$ and $w_W$ are adjusted so that the global width $w$ remains constant.

\NI While $h$, $r$, and $m$ control the strength of structural regularities, we assess their impact on network dynamics using deterministic neurons with $\tanh$ activation and three dynamical measures: $F$, $C$, and $N$.

\NI The {\bf fluctuation} $F\in[0,1]$ is the standard deviation of neuron outputs, averaged over neurons and time. It quantifies spontaneous activity; excessively large $F$ would hinder reliable computation.

\NI The {\bf correlation} $C\in[-1,+1]$ is defined as the product $y_m(t)\cdot y_n(t\!+\!1)$, averaged over neuron pairs and time. In a globally oscillatory state, $C\approx -1$, in a chaotic state $C\approx 0$, and in a fixpoint state $C\approx +1$. 

\NI Finally, the {\bf nonlinearity} $N\in[-1,+1]$ is the average operating regime of neurons. At $N\approx -1$, neurons operate linearly; at $N\approx +1$, they are saturated and behave digitally.

\NI While the primary focus of this study lies on how organizational regularities shape the intrinsic dynamics of RNNs, it is natural to ask whether these structural features also affect the network’s information processing performance. To address this, we embed the RNN into a reservoir computing framework and examine how the {\bf accuracy} $A$ in simple benchmark tasks varies as a function of the regularity control parameters.

\NI The following sections describe the construction of the weight matrices, the simulation setup for measuring dynamical indicators, and the test tasks used to evaluate computational performance.

% *******************************************************
\section{Methods}
% *******************************************************

\subsection{General Simulation Setup}

\NI The overall workflow of our investigation consists of the following steps:

\NI First, we {\bf set the control parameters}, including the distribution parameters—width $w$, density $d$, and excitatory-inhibitory balance $b$—as well as the regularity parameters: Dale homogeneity $h$, Hopfield reciprocity $r$, and modularity $m$.

\NI Second, we {\bf generate a set of $N_R$ random square weight matrices} that satisfy these specifications. Their properties are verified by empirically evaluating their statistical features.

\NI Third, for each matrix, we {\bf simulate the spontaneous dynamics of the corresponding RNN}. The network is initialized in a random state and then run freely for a large number of time steps. We refer to the resulting time series of neural activations as the output stream.

\NI Fourth, we {\bf compute the dynamical measures}—fluctuation $F$, correlations $C_0$ and $C_1$, and nonlinearity $N$—for each output stream, resulting in $N_R$ samples per measure. For visualization, we average these values across the ensemble of $N_R$ networks.

\NI In addition to analyzing intrinsic dynamics, we also {\bf examine how the regularity parameters affect information processing}. For this, the RNN is embedded into a reservoir computing framework: it receives input via a fixed input matrix, and its output stream is passed to a trainable readout layer. However, in the first part of the study, we set the input matrix to zero, so the RNN runs autonomously without external input, and the readout is not used. Nonetheless, we describe the full reservoir architecture in the following for completeness.

\subsection{Design of Reservoir Computer (RC)}

\NI The RC consists of an input layer, a recurrent reservoir, and a readout layer. The input data comprises $E$ consecutive episodes, each corresponding, for example, to a pattern to be classified. This data stream is fed into the reservoir and circulates through the system while propagating toward the output.

\NI At each time step $t$, the input layer receives $M$ parallel signals $x_m^{(t)} \in [-1, +1]$. These are linearly transformed by the input matrix $\mathbf{I}$ of size $N \times M$ and injected into the reservoir, as described by Eq.~\ref{yEq}. Each input episode spans $T$ time steps.

\NI The {\bf input layer} consists solely of the matrix $\mathbf{I}$ and is therefore purely linear. Its elements $I_{mn}$ are drawn independently from a normal distribution with zero mean and standard deviation $w_I$. To study autonomous RNN dynamics, we set $w_I = 0$, effectively decoupling the reservoir from external input.

\NI The {\bf reservoir} comprises $N$ recurrently connected neurons with $\tanh$ activation. At each time step, all neuron states $y_n$ are updated in parallel. Each neuron receives a bias term $b_{w,n}$, input from the external signals via $\mathbf{I}$, and recurrent input via the weight matrix $\mathbf{W}$ (see Eq.~\ref{yEq}). Initial states $y_n^{(0)}$ are drawn uniformly from $[-1, +1]$ and kept fixed for repeated simulations of the same reservoir. Different weight matrices receive independent initial states.

\NI The {\bf readout layer} performs an affine-linear transformation of the reservoir states $y_n$ using a $K \times N$ output matrix $\mathbf{O}$ and a bias vector $\mathbf{b_o}$, as in Eq.~\ref{zEq}. These parameters are trained via the pseudoinverse method (see below).

\NI The resulting outputs $z_k$ represent soft class votes. For classification, we apply the argmax function to obtain a discrete predicted class label $c$. This step introduces a nonlinearity that sharpens class boundaries in the output space.

\NI In summary, the RC is governed by the following equations:

\begin{eqnarray}
y_n^{(t)} & = & \tanh \left( 
  b_{w,n} + \sum_m I_{nm} x_m^{(t-1)} + 
  \sum_{n^{\prime}} W_{nn^{\prime}} y_{n^{\prime}}^{(t-1)} 
\right) \label{yEq} \\
z_k^{(t)} & = & b_{o,k} + \sum_n O_{kn} y_n^{(t)} \label{zEq} \\
c^{(t)} & = & \operatorname*{arg\,max} \left\{ z_k^{(t)} \right\}
\end{eqnarray}

\NI Although our main focus lies on classification tasks, we also include one regression-like task with continuous output, for which the argmax step is omitted.

\subsection{Sequence Generation Task}

\NI In general, we treat the reservoir computer as a trainable mapping from a sequence \( X \) of input vectors to a sequence \( Z \) of output vectors, where all vector components are real numbers constrained to a fixed range:
\[
X \in \left[-1,+1\right]^{TI \times M} \;\longrightarrow\; Z \in \left[-1,+1\right]^{TO \times K}
\]
Here, \( TI \) and \( TO \) denote the temporal lengths of the input and output sequences, while \( M \) and \( K \) are their respective vector dimensions.

\NI In the sequence generation task, the RNN is used as a deterministic dynamical system. At the start of each episode, an input sequence \( X \), randomly drawn from a set of \( N_{DC} \) discrete classes, drives the reservoir into a high-dimensional, class-specific 'priming state'. From the moment the input ends, the reservoir evolves autonomously through a sequence of internal states \( Y \), which the linear readout layer is trained to map onto a corresponding target sequence \( Z \).

\NI Importantly, the exact form of the internal state sequence \( Y \) is not critical, as long as it is high-dimensional and non-cyclic. The linear readout can typically learn to map such trajectories onto any desired non-cyclic low-dimensional output.

\NI What matters is that the reservoir consistently produces the same trajectory \( Y_c \) when primed with the same input sequence \( X_c \). This consistency breaks down if the reservoir exhibits unpredictable spontaneous activity or retains residual state information from previous episodes. The task thus provides a meaningful test of whether the reservoir’s dynamics is primarily driven by the current input—a necessary condition for reliable computation.

\NI To allow rapid evaluation across the parameter space and support two-dimensional phase diagrams with sufficient statistics, we use minimal task parameters: \( M = K = 2 \), \( TI = 1 \), \( TO = 2 \), and \( N_{DC} = 2 \). 

\NI The \( N_{DC} \) class-specific priming sequences \( X_c \) and their corresponding target sequences \( Z_c \) are defined prior to constructing the training and test datasets. All vector components are drawn independently from a uniform distribution over the interval \([-1, +1]\). In each episode, one of the \( N_{DC} \) cases is selected at random.

\subsection{Optimal Readout Layer Using Pseudoinverse}

\NI The weights and biases of the readout layer are computed using the pseudoinverse as follows:  

\NI Let \( Y \in \mathbb{R}^{(E-1) \times N} \) be the matrix of reservoir states directly after each input episode, where \( E \) is the total number of episodes and \( N \) is the number of reservoir neurons. Let \( Z \in \mathbb{R}^{(E-1) \times K} \) be the matrix of target output states, where \( K \) is the number of output units.  

\NI To account for biases in the readout layer, a column of ones is appended to \( Y \), resulting in the matrix \( Y_{\text{bias}} \in \mathbb{R}^{(E-1) \times (N+1)} \):  
\[
Y_{\text{bias}} = \begin{bmatrix} Y & \mathbf{1}_{E-1} \end{bmatrix}
\]
where \( \mathbf{1}_{E-1} \in \mathbb{R}^{(E-1) \times 1} \) is a column vector of ones.  

\NI The weights and biases of the readout layer are computed by solving the following equation using the pseudoinverse of \( Y_{\text{bias}} \):  
\begin{equation}
W_{\text{bias}} = Y_{\text{bias}}^+ Z
\label{optWei}
\end{equation}
where \( Y_{\text{bias}}^+ \) is the Moore-Penrose pseudoinverse of \( Y_{\text{bias}} \), and \( W_{\text{bias}} \in \mathbb{R}^{(N+1) \times K} \) contains both the readout weights and the biases.  

\NI To compute the pseudoinverse, we first perform a singular value decomposition (SVD) of \( Y_{\text{bias}} \):  
\[
Y_{\text{bias}} = U S V^\top
\]
where \( U \in \mathbb{R}^{(E-1) \times (E-1)} \) is a unitary matrix, \( S \in \mathbb{R}^{(E-1) \times (N+1)} \) is a diagonal matrix containing the singular values, and \( V^\top \in \mathbb{R}^{(N+1) \times (N+1)} \) is the transpose of a unitary matrix.  

\NI The pseudoinverse of \( Y_{\text{bias}} \) is computed as:  
\[
Y_{\text{bias}}^+ = V S^+ U^\top
\]
where \( S^+ \in \mathbb{R}^{(N+1) \times (E-1)} \) is the pseudoinverse of the diagonal matrix \( S \). The pseudoinverse \( S^+ \) is obtained by taking the reciprocal of all non-zero singular values in \( S \) and leaving zeros unchanged.  

\NI Finally, after inserting \( Y_{\text{bias}}^+ \) into Eq.~\ref{optWei}, the optimal readout weights \( W \in \mathbb{R}^{K \times N} \) and biases \( b \in \mathbb{R}^K \) are extracted from the extended matrix \( W_{\text{bias}} \) as  
\[
W = \left( W_{\text{bias}}^\top \right)_{1:N,\,:}
\]
\[
b_w = \left( W_{\text{bias}}^\top \right)_{N+1,\,:}
\]
where the first \( N \) rows of \( W_{\text{bias}}^\top \) define the readout weights and the last row defines the biases.

\subsection{Generation of Weight Matrices with Homogeneity, Reciprocity, or Modularity}

\NI The generation of weight matrices is based on the six control parameters defined in the Introduction. Depending on the selected values of homogeneity $h$, reciprocity $r$, and modularity $m$, different structural features are incorporated into otherwise random matrices.

\NI We begin by generating a matrix of magnitudes $m_{ij}$, where each entry is drawn from a normal distribution with zero mean and standard deviation $w$, then made positive via $m_{ij} := |m_{ij}|$.

\NI A binary mask matrix $n_{ij} \in \{0,1\}$ is created, where each element is set to $1$ with probability $d$. This defines the sparsity structure.

\NI A sign matrix $s_{ij} \in \{-1, +1\}$ is generated, with $+1$ assigned with probability $(b + 1)/2$ to control the excitatory-inhibitory balance.

\NI The elementwise product of magnitude, mask, and sign matrices yields the pure weight matrix $W^{(pure)} = m_{ij} \cdot n_{ij} \cdot s_{ij}$, which serves as the default base.

\NI To implement homogeneity ($h > 0$, $r = 0$), we generate a Dale-conform matrix $W^{(Dale)}$ in which each column (corresponding to one sending neuron) has a uniform sign. The final weight matrix $W$ is then obtained by stochastic interpolation: $w_{ij} := w^{(Dale)}_{ij}$ with probability $h$, and $w^{(pure)}_{ij}$ with probability $1 - h$.

\NI The Dale matrix uses the same magnitudes $m_{ij}$ and mask $n_{ij}$ as $W^{(pure)}$, but assigns each column a fixed sign (either all $+1$ or all $-1$), again with $(b + 1)/2$ controlling the excitatory fraction.

\NI To implement reciprocity ($r > 0$, $h = 0$), we generate a symmetric Hopfield-like matrix $W^{(Hopf)}$ by copying the upper triangle and diagonal of $W^{(pure)}$ into the lower triangle. The final matrix is again obtained via stochastic interpolation: $w_{ij} := w^{(Hopf)}_{ij}$ with probability $r$, and $w^{(pure)}_{ij}$ otherwise.

\NI To implement modularity ($m > 0$), we divide the $N \times N$ matrix into regular square blocks of size $S \times S$, grouping the $N$ neurons into $N/S$ modules. Each block is randomly assigned to the 'strong' or 'weak' class, with the probability of 'strong' given by $f_{SB}$. The realized fractions $f_s$ and $f_w = 1 - f_s$ are determined empirically after assignment.

\NI Weak blocks are filled with Gaussian values of standard deviation $w \cdot q_w$, where $q_w = 1 - m$. Strong blocks use standard deviation $w \cdot q_s$, with
\[
q_s = \sqrt{ \frac{1 - f_w q_w^2 }{f_s} } = \sqrt{ \frac{1 - f_w (1 - m)^2 }{1 - f_w} }
\]
to ensure that the total matrix standard deviation remains $w$.

\NI For $m = 0$, both scaling factors are equal to one; for $m = 1$, the weak blocks vanish and $q_s = \sqrt{1 / f_s}$.

\NI Except for the extreme case $m = 1$, where all weak blocks vanish, the resulting matrix is fully dense ($d = 1$). To impose the desired balance $b$, all elements are made positive and then flipped to negative with probability $(1 - b)/2$.

\NI For clarity, we never apply homogeneity ($h$), reciprocity ($r$) and modularity ($m$) simultaneously.

\subsection{Evaluation of Weight Matrices}

\NI In addition to the control parameters $d, b, h, r$, we define corresponding {\bf empirical parameters} $D, B, H, R$ that quantify the actual density, balance, homogeneity, and reciprocity of a given weight matrix $\mathbf{W}$.

\NI The empirical {\bf density} $D$ is computed as the fraction of non-zero elements:
\[
D = \frac{n_{\text{nonzero}}}{n_{\text{total}}}, \quad \text{where } n_{\text{nonzero}} = \#\{w_{ij} \ne 0\}
\]

\NI The empirical {\bf balance} $B$ measures the ratio of excitatory to inhibitory weights:
\[
B = \frac{n_{\text{pos}} - n_{\text{neg}}}{n_{\text{pos}} + n_{\text{neg}}}, \quad \text{with } n_{\text{pos}} = \#\{w_{ij} > 0\}, \; n_{\text{neg}} = \#\{w_{ij} < 0\}
\]

\NI The empirical {\bf homogeneity} $H$ is computed column-wise. For each column $c$, we calculate
\[
H_c = \frac{|n_{\text{pos}} - n_{\text{neg}}|}{n_{\text{pos}} + n_{\text{neg}}}
\]
where $n_{\text{pos}}$ and $n_{\text{neg}}$ refer to the number of positive and negative entries in column $c$. If all elements in the column are zero, we set $H_c := 0$. The global homogeneity is the mean over all columns:
\[
H = \langle H_c \rangle_c
\]

\NI The empirical {\bf reciprocity} $R$ is computed from all off-diagonal pairs $(i, j)$ in the upper triangle:
\[
R_{ij} = 1 - \frac{|w_{ij} - w_{ji}|}{|w_{ij}| + |w_{ji}|}
\]
If both $w_{ij}$ and $w_{ji}$ are zero, we define $R_{ij} := 1$. The matrix-level reciprocity is the average over all such pairs:
\[
R = \langle R_{ij} \rangle_{i < j}
\]

\NI For each set of control parameters $(d, b, h, r)$, we generate an ensemble of weight matrices and compute the empirical parameters $D, B, H, R$ as averages over the corresponding values from each individual matrix.

\subsection{Fluctuation Measure}

\NI The neural fluctuation measure \( F \) quantifies the average temporal variability of reservoir activations. For each neuron \( n \), we compute the standard deviation \( \sigma_n \) of its activation time series \( y_n^{(t)} \). The global fluctuation is defined as the mean over all neurons:
\[
F = \left\langle \sigma_n \right\rangle_n
\]

\NI Since $\tanh$-neurons produce outputs in \([-1, +1]\), the fluctuation \( F \) lies in \([0, 1]\). A value of \( F = 0 \) indicates a resting or fixpoint state, while \( F = 1 \) corresponds to perfect two-state oscillation (e.g., alternating between $+1$ and $-1$).

% ********************************************

\subsection{Correlation Measure}

\NI To assess temporal correlations, we compute the average product of the activation of neuron \( m \) at time \( t \) and neuron \( n \) at time \( t + \Delta t \):
\[
C_{mn}^{(\Delta t)} = \left\langle y_m^{(t)} \cdot y_n^{(t\!+\!\Delta t)} \right\rangle_t
\]

\NI Unlike the Pearson correlation coefficient, we deliberately avoid subtracting the mean or normalizing by the standard deviations. This ensures that the matrix elements \( C_{mn}^{(\Delta t)} \) remain well-defined even when one or both signals are constant, as in a fixpoint state.

\NI The global correlation measure is defined as the average over all neuron pairs:
\[
C_{\Delta t} = \left\langle C_{mn}^{(\Delta t)} \right\rangle_{mn}
\]

\NI Owing to the bounded output of the $\tanh$ neurons, the correlation values \( C_{\Delta t} \) always lie within the range \([-1, +1]\).

% ********************************************

\subsection{Nonlinearity Measure}

\NI The shape of the activation distribution \( p(y) \) reflects whether the reservoir operates in a linear or nonlinear regime. A central peak at \( y = 0 \) indicates a linear regime; two peaks near \( \pm1 \) indicate saturation and thus nonlinearity.

\NI We define a nonlinearity measure
\[
\alpha = f_A - f_B + f_C
\]
based on the fractions of neural activations falling into the following intervals:
\[
\begin{array}{ll}
f_A & \in [-1,\;-0.5) \\
f_B & \in [-0.5,\;+0.5] \\
f_C & \in (+0.5,\;+1]
\end{array}
\]

\NI The resulting measure \( \alpha \in [-1, +1] \) distinguishes three regimes:  
$\alpha \approx -1$ for linear operation, $\alpha \approx 0$ for intermediate or flat activation, and $\alpha \approx +1$ for saturated, digital-like behavior.

\NI This intuitive yet robust definition proved most effective among several tested alternatives. It captures the essential qualitative transition in \( p(y) \) from unimodal (linear) to bimodal (nonlinear) distributions, as highlighted in earlier studies.

% ********************************************

\subsection{Accuracy Measure}

\NI We evaluate performance by comparing the actual output sequences \( Z_{\text{act}} \) of the readout layer with the corresponding target sequences \( Z_{\text{tar}} \), and compute the root-mean-square error \( E_{\text{RMS}} \). This error is normalized by the standard deviation \( \Delta Z_{\text{tar}} \) of the target sequences.

\NI To obtain an accuracy measure \( A \in [0, 1] \), we define:
\[
A = \frac{1}{1 + (E_{\text{RMS}} / \Delta Z_{\text{tar}})}
\]

\NI Note that \( A \approx 0.5 \) when the RMS error is comparable to the variability of the target data, and \( A = 1 \) when the output \( Z_{\text{act}} \) matches the target \( Z_{\text{tar}} \) exactly.

% *******************************************************
\section{Results}

% *******************************************************

\subsection{Validation of Control Parameters}

\NI As described in the Methods section, we generate weight matrices with prescribed values for connection density \( d \), excitatory/inhibitory balance \( b \), Hopfield reciprocity \( r \), Dale homogeneity \( h \), and modularity \( m \). To verify that the generation procedures operate as intended, we compute the corresponding empirical parameters directly from the generated matrices. Throughout this paper, lowercase letters \( d, b, r, h, m \) denote the prescribed control parameters, while uppercase letters \( D, B, R, H \) refer to their empirically measured counterparts. Validation of the modular structure—additionally dependent on block size \( S \) and the fraction \( f_{SB} \) of 'strong' blocks—is addressed separately (see below).

\begin{figure}[ht!]
%\centering
\includegraphics[width=0.95\linewidth]{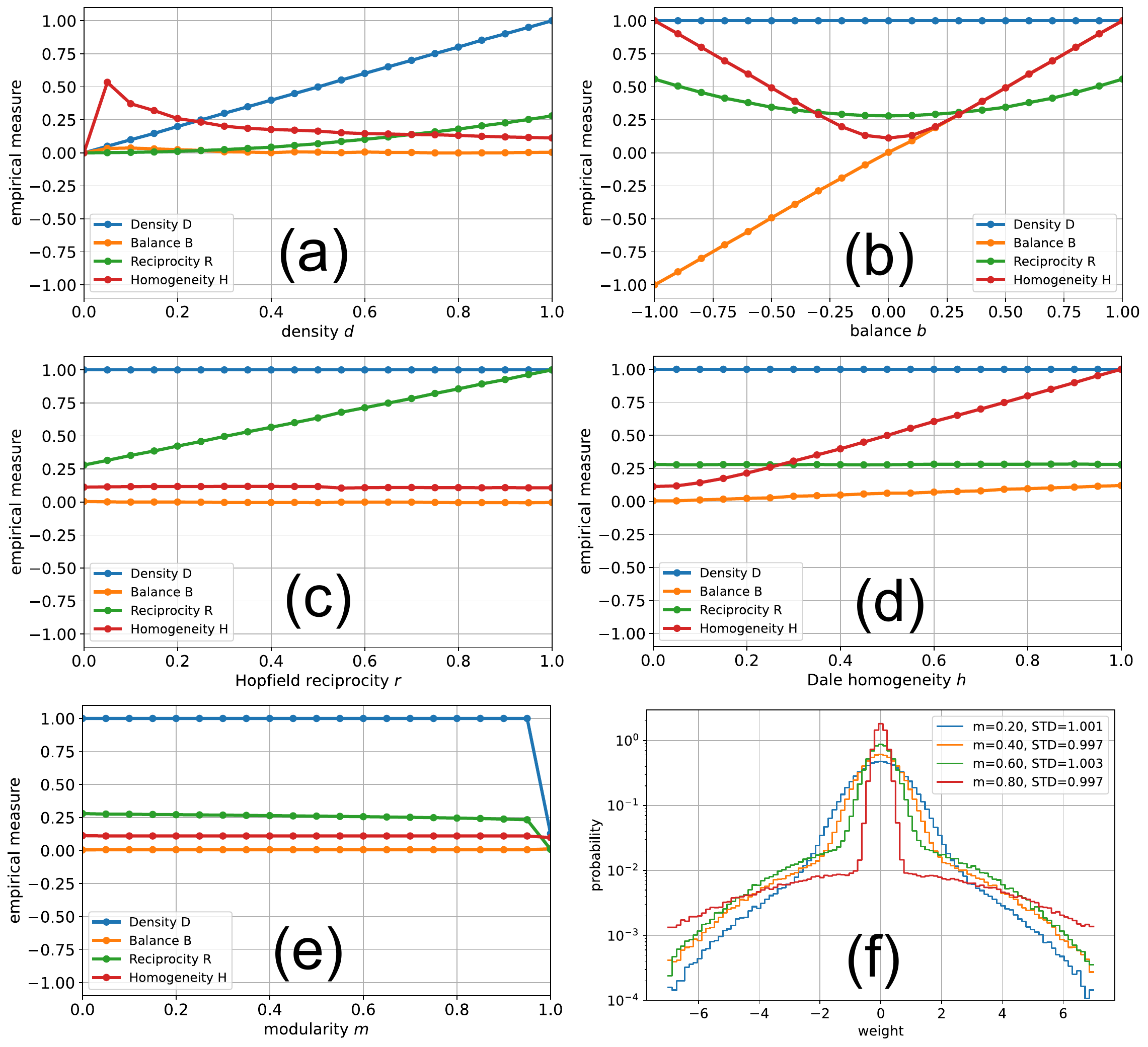}
\caption{{\bf Prescribed and empirical control parameters}. We use a weight matrix of size $50\times50$ with parameters initially set to standard values $d=1$ and $b=r=h=m=0$. In each of the plots (a-e), one control parameter $x$ is scanned through its full permissible range, while all others remain at their standard values. The empirical measures $D,B,R,H$ are evaluated as a function of the scanned parameter $x$. \textbullet$\;${\bf(a)} Scan of the density $d$. \textbullet$\;${\bf(b)} Scan of the balance $b$. 
\textbullet$\;${\bf(c)} Scan of the reciprocity $r$. \textbullet$\;${\bf(d)} Scan of the homogeneity $h$.
\textbullet$\;${\bf(e)} Scan of the modularity $m$.
\textbullet$\;${\bf(f)} Probability distributions of weight matrix elements for different degrees of modularity $m$, using a $1000\times1000$ matrix with $w=d=1$ and $b=0$. Block size was $S=100$, with a fraction of strong blocks $f_{SB}=0.1$. Inset shows that the standard deviation of the distributions remains constant.
} 
\label{fig_PreEmp}
\end{figure}

\NI For validation, we first fix all control parameters \( d, b, r, h, m \) to standard values. Then, one parameter \( x \) is varied across its entire admissible range while the others remain fixed. During this one-dimensional scan, we compute all empirical quantities \( D, B, R, H \) as functions of \( x \) (Fig.~\ref{fig_PreEmp}(a–e)). Ideally, each control parameter \( x \) should primarily affect its corresponding empirical statistic \( X \), without significantly altering the others. However, certain interdependencies are inevitable due to shared structural properties.

\NI As shown in panel (a), the prescribed density \( d \) directly controls the empirical density \( D \), with negligible effect on the balance \( B \). However, it also influences reciprocity \( R \) and homogeneity \( H \) to a minor extent.

\NI Similarly, the prescribed balance \( b \) determines the empirical balance \( B \), while leaving \( D \) unaffected (panel b). Nonetheless, \( R \) and \( H \) increase as the system becomes more unbalanced. Even for perfectly balanced matrices at \( b = 0 \), there exists a minimal, unavoidable level of reciprocity and homogeneity due to the Gaussian weight distribution.

\NI Varying the prescribed Hopfield reciprocity \( r \) results in a nearly linear increase of the empirical reciprocity \( R \), while the other measures remain unaffected (panel c).

\NI Increasing the prescribed Dale homogeneity \( h \) leads to a monotonic rise in empirical homogeneity \( H \), along with a slight increase in balance \( B \). Reciprocity \( R \) and density \( D \) remain essentially unchanged (panel d).

\NI Finally, the prescribed modularity \( m \) has virtually no effect on any of the empirical measures \( D, B, R, H \), except at the extreme value \( m = 1 \)(panel e).

\NI To validate the modularity construction, we consider a \( 1000 \times 1000 \) weight matrix with parameters \( w = d = 1 \) and \( b = 0 \). The block size is set to \( S = 100 \), and the fraction of strong blocks to \( f_{SB} = 0.1 \). We then gradually increase the modularity parameter \( m \) from 0.2 to 0.8, and compute the histogram of matrix elements for each value (panel f). 

\NI A semi-logarithmic plot reveals that the resulting distributions are mixtures of two Gaussians with distinct standard deviations. As expected, the mixture preserves the global distribution width (STD), as seen in the insets of panel (f).

\subsection{RNN Phase Diagrams without Input}

\begin{figure}[ht!]
%\centering
\includegraphics[width=0.95\linewidth]{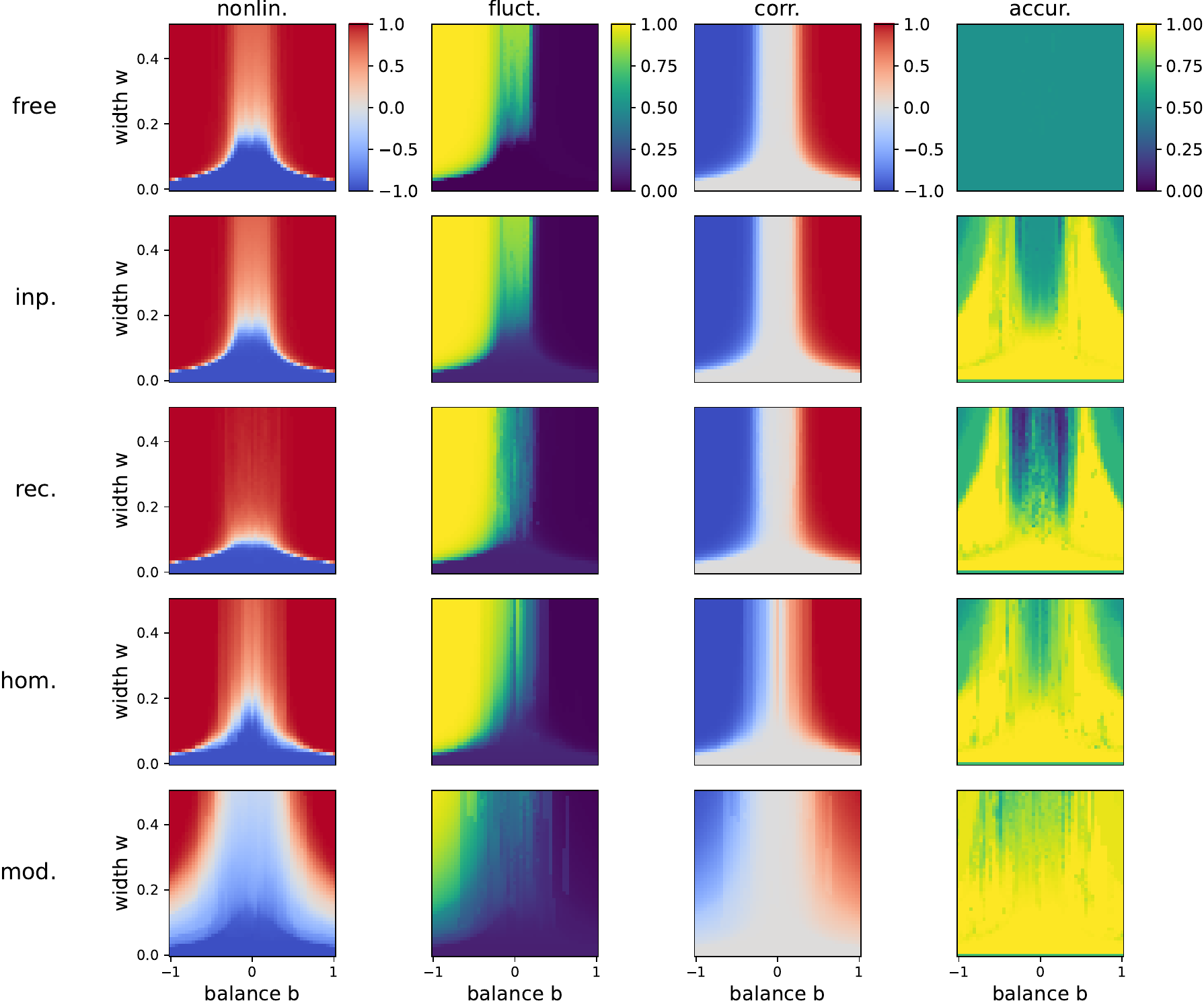}
\caption{{\bf Phase diagrams of RNN dynamics and computational performance, as functions of the balance $b$ and the width $w$}. In all cases, the RNN consists of 50 arctan-neurons, fully connected (density $d=1$). \textbullet$\;$ The four columns of phase diagrams, from left to right, correspond to the {\bf nonlinearity} $N$, the {\bf fluctuation} $F$, the {\bf correlation} $C_1$, and the {\bf accuracy} $A$ in a sequence generation task (For details see Methods and Results). \textbullet$\;$ The top row corresponds to a free-running (no input) RNN, with all regularity parameters set to zero.
All plots below correspond to RNNs driven by inputs and used for computations. 2nd row: All regularity parameters set to zero. 3rd row: Hopfield reciprocity parameter set to $r=0.9$. 4th row: Dale homogeneity parameter set to $h=0.9$. 5th row: modularity parameter set to $m=0.9$. \textbullet$\;$ Note that the phase diagram of the nonlinearity parameter in the free-running system (upper left) shows four main dynamical regions characterized by quiescence (Q, blue dome at the bottom), chaos (C, pale red stripe around the upper center), oscillations (O, red flank at the left side) and fixpoints (F, red flank at the right side). Turning on the regularity parameters $r$, $h$ and $m$ has a clear effect on the dynamical variables $N$, $F$, $C_1$, as well on the accuracy $A$.
} 
\label{fig_PD}
\end{figure}

\NI In the following section, we consider an RNN of 50 neurons, randomly linked to each other according to a weight matrix of full connection density $d=1$. We analyze how the dynamical and information processing properties of the RNN change as a function of the excitatory-inhibitory balance $b$ and the width $w$ of the Gaussian distribution of connection strengths. The results are presented in the form of 'phase diagrams', where the dynamical measures $N$ (nonlinearity), $F$ (fluctuation), $C$ (correlation at time lag $\Delta t=1$), and $A$ (accuracy in a computational task) are color-coded in the $b$-$w$ plane (compare Fig.\ref{fig_PD}).

\NI In the free-running RNN (upper row in Fig.\ref{fig_PD}), we identify four characteristic dynamical regimes, most clearly visible in the nonlinearity phase diagram $N(b,w)$:

\NI In the lower central part of the phase plane lies the {\bf quiescent region $\mathsf{Q}$}, roughly shaped like 'Mt Fuji'. In this region, at $w\approx 0$, the neurons are virtually unconnected. Therefore all activations remain at very small values, determined only by the individual biases. As indicated by the nonlinearity measure $N\approx -1$, the neurons operate here in the linear regime, near the center of the tanh activation function. The temporal constancy of the activations is reflected in the fluctuation measure $F\approx 0$. While the Pearson correlation coefficient would diverge for constant zero signals, our non-normalized correlation measure simply yields $C\approx 0$ in this case.

\NI In the right wing of the phase plane lies the {\bf fixpoint regime $\mathsf{F}$}. Also marked by temporally constant activations, it exhibits fluctuations $F\approx 0$. However, due to strong coupling strengths ($w>0$) and predominantly excitatory weights ($b>0$), a positive feedback loop drives the network into high-activation global fixpoints, where each neuron becomes trapped in either the positive or negative saturation of the activation function. As a result, the nonlinearity measure $N\approx +1$ indicates that the neurons now operate in a digital regime, and the correlation measure yields $C\approx +1$, showing that the neurons retain the same digital value over time.

\NI In the left wing of the phase plane lies the {\bf oscillatory regime $\mathsf{O}$}. Here, strong ($w>0$), predominantly inhibitory ($b<0$) coupling leads to global periodic flips of all neurons between the two saturation states. Consequently, the correlation $C\approx -1$ reflects activation values of opposite sign between successive time steps. The fluctuation $F\approx 1$ indicates high-amplitude temporal variation, and the nonlinearity $N\approx +1$ shows that the neurons operate digitally.

\NI In the upper central part of the phase plane lies the {\bf chaotic regime $\mathsf{C}$}. The dynamics here are also driven by strong mutual couplings ($w>0$), but now the approximate balance between excitatory and inhibitory connections ($b\approx 0$) leads to much more complex and irregular behavior—both over time and across neurons. This is reflected in vanishing correlations $C\approx 0$. The neurons operate in a mixed linear, intermediate, or digital regime, such that the nonlinearity measure $N$ lies somewhere between $-1$ and $+1$. The corresponding temporal fluctuations $F$, on average, are smaller than in the oscillatory regime.

\subsection{RNN Phase Diagrams with Input}

\NI Next, while the RNN is continuously updating, we feed in two time-dependent input signals related to a computational task. For this purpose, we use a dense $50\times 2$ input matrix $\mathbf{I}$, so that every neuron receives both inputs. The elements of $\mathbf{I}$ are drawn from a Gaussian distribution with zero mean and a standard deviation of 0.3. The two input signals range between $-1$ and $+1$, and all further details are described in the Methods section.

\NI We find that the injection of inputs has only a very weak effect on the phase diagrams of $N$, $F$, and $C$ (see second row in Fig.\ref{fig_PD}). Only the fluctuation level $F$ in the quiescent regime $\mathsf{Q}$ is slightly elevated compared to the input-free case, which is to be expected.

\NI We also add a readout layer, optimized by the method of the pseudo-inverse, which transforms the global time-dependent states of the RNN into output signals. In our case, the readout layer is optimized for a sequence generation task (see Methods section for details), and the performance of the resulting reservoir computer is measured by an accuracy value $A$ that ranges between zero and one (right-most phase diagram in row 2).

\NI We find that the accuracy remains close to one throughout the entire quiescent regime $\mathsf{Q}$. As shown in a previous publication, this regime is well-suited for many types of tasks, as the RNN states are then primarily determined by the input, not by spontaneous internal dynamics.

\NI The accuracy drops considerably within the chaotic regime $\mathsf{C}$, where the irregular and unpredictable dynamics of the RNN interfere with the execution of the computational task.

\NI A strong reduction in accuracy is also observed in the upper and outer parts of the oscillatory ($\mathsf{O}$) and fixpoint ($\mathsf{F}$) regimes. There, the autonomous RNN dynamics are predictable, but the neurons are driven so deeply into the saturation regime that task-related computations can no longer be carried out.

\NI Remarkably, the accuracy remains close to one in the narrow regions between the chaotic regime $\mathsf{C}$ and the neighboring oscillatory $\mathsf{O}$ or fixpoint $\mathsf{F}$ regimes. These two 'edges of chaos' thus prove suitable for task-related computation; however, they become increasingly narrow as the recurrent coupling strength $w$ is increased.

\subsection{Effect of strong Hopfield Reciprocity}

\NI Next, we leave the reservoir computer unchanged, with the only modification being the introduction of a relatively strong degree of Hopfield reciprocity, $r=0.9$, into the RNN's weight matrix (see third row in Fig.\ref{fig_PD}).

\NI Comparing the nonlinearity phase diagram at $r=0.9$ with the corresponding diagram at $r=0$, we observe a significant shrinking of the quiescent regime $\mathsf{Q}$ and an increase in nonlinearity within the chaotic regime $\mathsf{C}$.

\NI At the same time, a comparison of the fluctuation diagrams $F(b,w\;|\;r\!=\!0)$ and $F(b,w\;|\;r\!=\!0.9)$ reveals that the fluctuation amplitudes of the neural activations in the chaotic regime $\mathsf{C}$ are reduced, on average, by the introduction of Hopfield reciprocity. Combined with the higher nonlinearity, this suggests that a part of the neurons now remain in a fixed state and do not participate in the chaotic dynamics.

\NI No differences are observed between the correlation phase diagrams $C(b,w\;|\;r\!=\!0)$ and $C(b,w\;|\;r\!=\!0.9)$. 

\NI Finally, the accuracy phase diagram $A(b,w\;|\;r\!=\!0.9)$ shows that performance in the chaotic regime has dropped to levels even below those of the $r=0$ system. This suggests that Hopfield reciprocity can, at least in certain computational tasks, have a detrimental effect on the performance of a reservoir computer.

\subsection{Effect of strong Dale Homogeneity}

\NI Starting again from the standard case without any structural regularities ($r = h = m = 0$), we now set the Dale homogeneity parameter to $h = 0.9$ and recompute the four phase diagrams (see fourth row in Fig.~\ref{fig_PD}).

\NI The diagrams for nonlinearity, fluctuation, and correlation at $h = 0.9$ resemble those observed at $r = 0.9$ more closely than the standard case.

\NI However, the accuracy diagram $A(b,w\,|\,h = 0.9)$ shows that Dale homogeneity improves task performance even beyond the standard case $A(b,w\,|\,r = h = m = 0)$. In particular, within the chaotic regime $\mathsf{C}$, the reservoir computer can now tolerate significantly higher coupling strengths $w$.

\subsection{Effect of strong Modularity}

\NI Finally, we set the modularity parameter to $m = 0.9$, using block sizes of $SNB = 10$ neurons and a strong block fraction of $f_{SB} = 0.1$ (see Methods for details). Modularity exerts a pronounced influence on all phase diagrams (see fifth row in Fig.~\ref{fig_PD}):

\NI The nonlinearity and correlation diagrams indicate that the oscillatory and fixpoint regimes now occupy only a narrow region of phase space, restricted to highly unbalanced weights $|b| \approx 1$ and strong coupling $w > 0.25$.

\NI Meanwhile, the previously chaotic regime is replaced by a broad region in which both nonlinearity and fluctuations remain moderate, while low correlation values still suggest irregular (non-repetitive) dynamics.

\NI Most strikingly, the accuracy now reaches very high levels across almost the entire phase diagram. This indicates that modularity renders even formerly unproductive regimes—oscillatory, fixpoint, and chaotic—computationally useful.

\subsection{Effect of Gradually Increasing Regularity}

\begin{figure}[ht!]
%\centering
\includegraphics[width=0.9\linewidth]{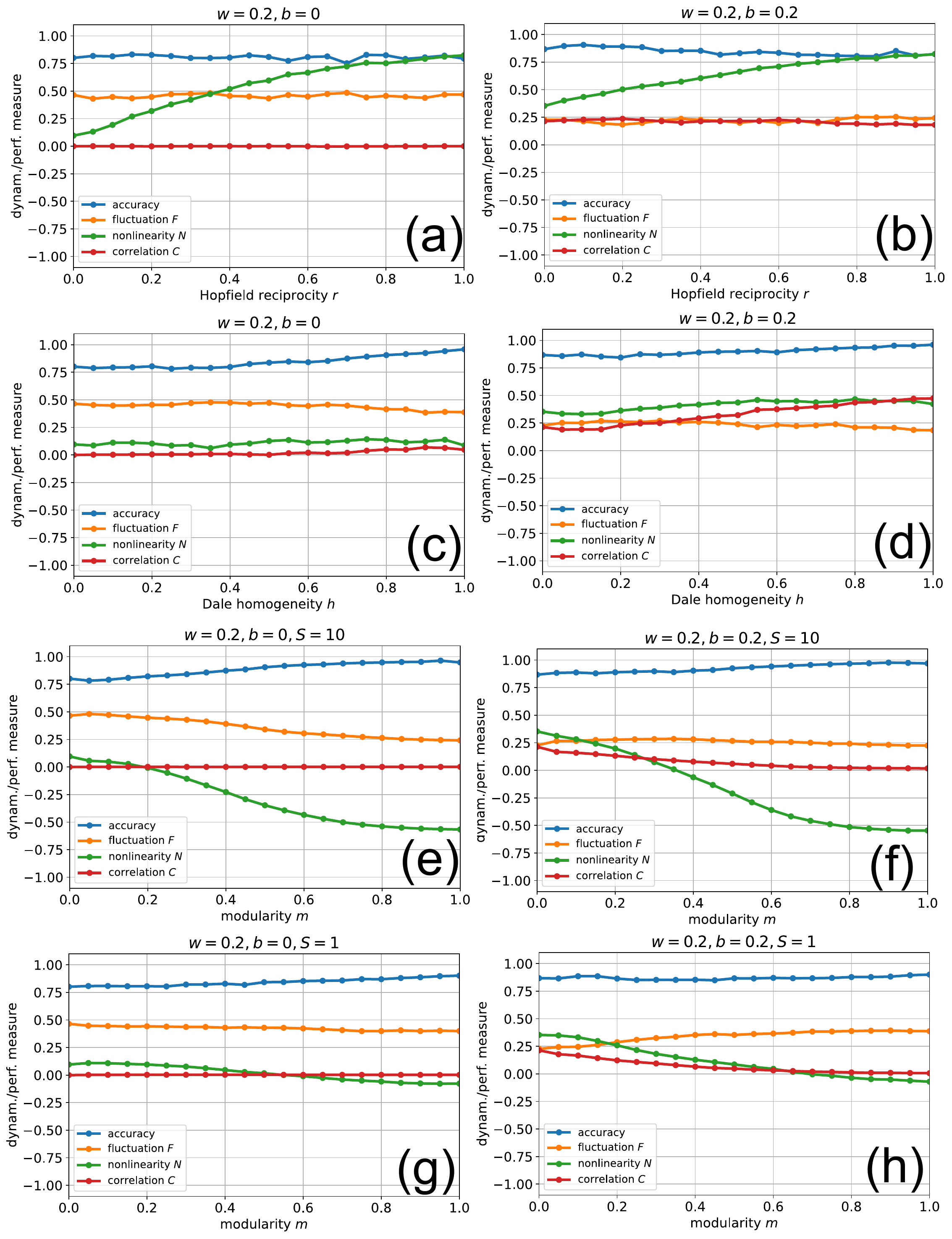}
\caption{{\bf Effect of increasing regularity on RNN dynamics and computational performance.} A RNN with 50 neurons and width $w=0.2$ is used in the sequence generation task. We compute the fluctuation $F$, the nonlinearity $N$, the correlation $C$ and the accuracy $A$ as one of the regularity parameters $r,h,m$ is scanned through its entire permissible range. The left column of plots is for balance $b=0$, the right for $b=0.2$. \textbullet$\;$ (a,b) Scan of Hopfield reciprocity $r$.
\textbullet$\;$ (c,d) Scan of Dale homogeneity $h$.
\textbullet$\;$ (e,f) Scan of modularity $m$ for block size $S=10$.
\textbullet$\;$ (g,h) Scan of modularity $m$ for block size $S=1$. The fraction of strong blocks was $f_{SB}=0.1$ in all modularity scans.
} 
\label{fig_Scan}
\end{figure}

\NI We now examine how a gradual increase of the organizational regularity parameters \( r, h, m \) affects the dynamical quantities \( N, F, C \) and the computational performance \( A \) (Fig.~\ref{fig_Scan}(a–h)). All simulations use an RNN with \( N = 50 \) neurons and connection width \( w = 0.2 \), comparing two balance settings: \( b = 0 \) (chaotic regime) and \( b = 0.2 \) (weak fixpoint regime). The RNN is used throughout as a reservoir in the sequence generation task. While one regularity parameter is varied over its full range \([0,1]\), the others are held at zero.

\NI Increasing Hopfield reciprocity \( r \) leads to a marked rise in nonlinearity \( N \) for both balance conditions, while fluctuation \( F \) and correlation \( C \) remain largely unaffected. In the fixpoint regime (\( b = 0.2 \)), reciprocity slightly degrades computational performance (panels a, b).

\NI Increasing Dale homogeneity \( h \) leads to distinct effects in both regimes. For \( b = 0 \), homogeneity reduces both fluctuation \( F \) and nonlinearity \( N \), resulting in a strong performance gain. For \( b = 0.2 \), \( h \) induces a rise in both correlation \( C \) and nonlinearity \( N \), yet performance still improves (panels c, d).

\NI Increasing modularity \( m \), with a block size \( S = 10 \), yields particularly substantial effects. In the chaotic regime (\( b = 0 \)), both fluctuation \( F \) and nonlinearity \( N \) are significantly reduced, leading to a notable performance boost. For \( b = 0.2 \), modularity lowers nonlinearity \( N \) and slightly reduces correlation \( C \); the resulting performance gain is weaker but still apparent (panels e, f).

\NI When the block size is reduced to \( S = 1 \), the weight matrix no longer contains distinct blocks. In this case, increasing \( m \) merely changes the weight distribution from a single Gaussian to a mixture with unchanged total standard deviation. For \( b = 0 \), this still yields a modest performance increase, though the dynamical changes are less pronounced compared to the block-structured case. For \( b = 0.2 \), no further improvement is observed (panels g, h).

\subsection{Effect of regularities on time-dependent neuron activations}

\begin{figure}[ht!]
%\centering
\includegraphics[width=0.9\linewidth]{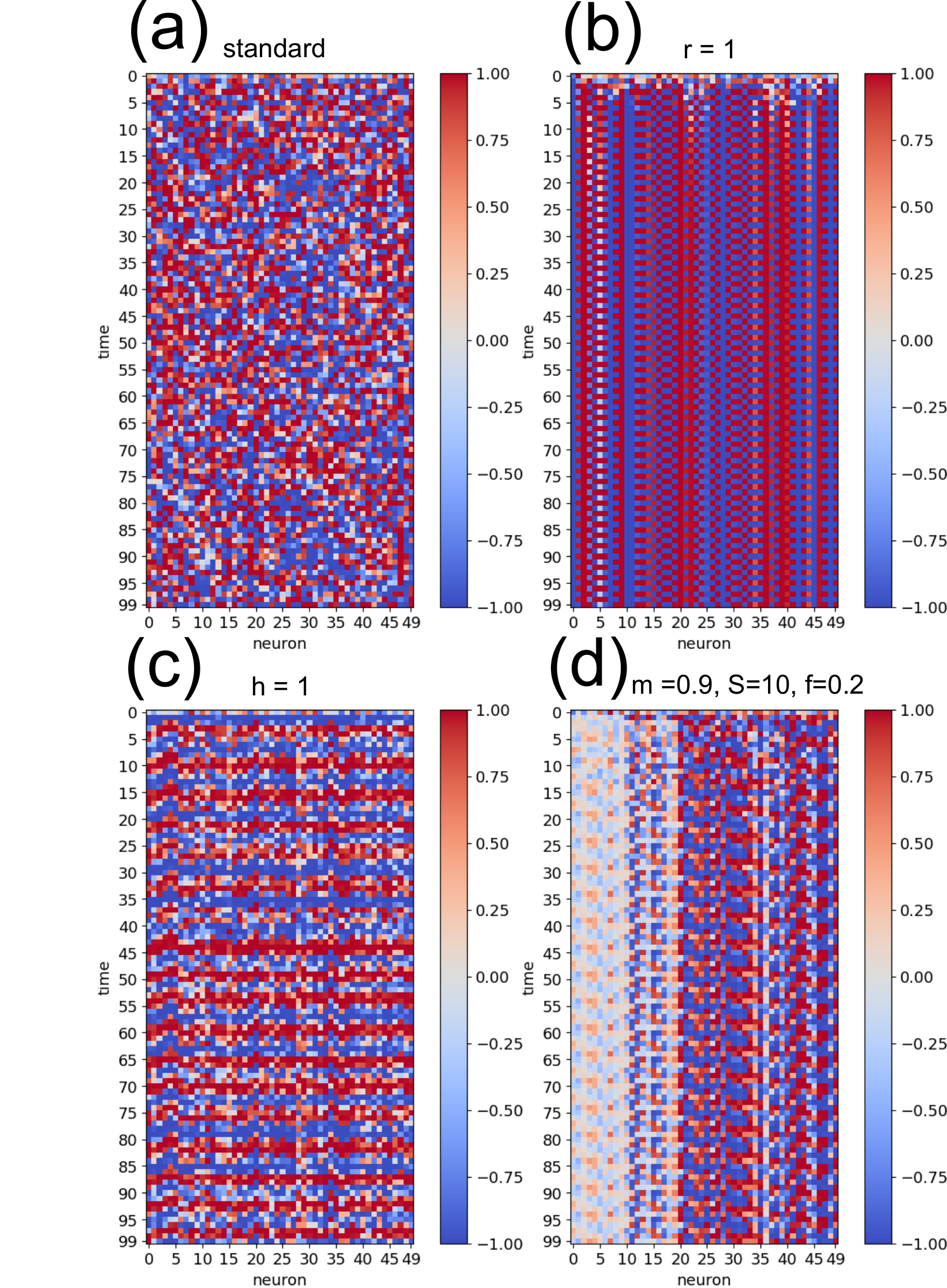}
\caption{{\bf Example neuron activations over time}. \textbullet$\;${\bf(a)} Standard, free-running RNN of 50 neurons, connected with density $d=1$, balance $b=0$, width $w=0.5$, and all organizational regularity parameters set to zero: $r=h=m=0$.
\textbullet$\;${\bf(b)} RNN with maximal Hopfield reciprocity $r=1$ and all other parameters unchanged.
\textbullet$\;${\bf(c)} RNN with maximal Dale homogeneity $h=1$.
\textbullet$\;${\bf(d)} RNN with modularity $m=0.9$, block size $S=10$, and probability $f_{SB}=0.2$.}
\label{fig_ActPlt}
\end{figure}

\NI To better understand how the regularity parameters influence RNN dynamics, we examine the time-dependent activations of neurons under four distinct parameter settings (Fig.~\ref{fig_ActPlt}(a–d)). Multiple runs were compared to ensure that the behaviors shown are representative.

\NI In the standard, free-running RNN with 50 neurons—fully connected ($d=1$), zero balance ($b=0$), weight width $w=0.5$, and all regularity parameters set to zero—the network operates in a chaotic regime (panel a). As expected, there are no apparent temporal correlations within single neurons (columns in the plot) or spatial correlations across neurons (rows).

\NI When Hopfield reciprocity is set to its maximum value $r=1$ (panel b), the dynamics become highly ordered. Some neurons maintain constant activations, while others exhibit period-two oscillations. Although the system is no longer chaotic, it does not settle into a global fixed point or simple periodic attractor.

\NI With Dale homogeneity set to its maximum value $h=1$ (panel c), we observe a weak form of instantaneous (spatial) synchronization across neurons, indicated by similar colors along rows. The dynamics are approximately periodic, with a period greater than two.

\NI With modularity set to $m=0.9$ (using $S=10$ and $f_{SB}=0.2$), the behavior clearly separates into distinct groups of neurons. Each group displays longer-period oscillations, with some groups showing weak and others strong amplitude fluctuations.

% *******************************************************
\section{Discussion}

\NI In this work, we investigated how three distinct organizational regularities—Hopfield reciprocity, Dale homogeneity, and modularity—affect the dynamical behavior and computational performance of recurrent neural networks (RNNs).

\subsection{Prior Expectations and Numerical Findings}

\NI  In the following, we interpret our numerical results in light of prior assumptions and expectations concerning each of these structural features.

\subsubsection*{Hopfield Reciprocity}

\NI Hopfield-type reciprocity introduces symmetry into the weight matrix by increasing the probability that any connection from neuron A to B is mirrored by an equal connection from B to A. In classical Hopfield networks, such symmetry is instrumental in stabilizing fixed-point attractors, corresponding to stored patterns. These networks use binary units and asynchronous updates, enabling the system to settle gradually into one of the memorized configurations. The symmetric weights ensure that the energy landscape has defined minima, guiding the dynamics toward stable fixed points.

\NI In our model, the situation differs in several respects. The neurons are continuous-valued with $\tanh$ activation, and all updates occur synchronously. Nonetheless, it was natural to expect that increasing Hopfield reciprocity \( r \) might exert an ordering influence on the network dynamics—possibly reducing chaotic fluctuations or promoting temporal coherence. It was unclear, however, whether such symmetry would be sufficient to generate global attractors under the more complex dynamics of our setup.

\NI The numerical findings only partially confirm this expectation. While the correlation and fluctuation measures remain nearly unaffected by increasing \( r \), we observe a significant rise in the nonlinearity parameter \( N \), particularly in balanced networks (\( b \approx 0 \)), which previously exhibited chaotic behavior. In this regime, the neurons increasingly operate in saturated mode (\( y \approx \pm 1 \)), indicating a shift toward quasi-digital dynamics.

\NI We interpret this phenomenon as a form of local stabilization: reciprocal neuron pairs mutually reinforce each other's output, pushing one another deeper into the same saturation branch of the activation function. This feedback does not generate coherent global attractors or fixed points, but it increases the likelihood that individual neurons—or small clusters—settle into persistent digital-like states. In essence, reciprocity acts as a local ordering force that fragments the system into semi-stable subdynamics.

\NI However, this type of local stabilization does not benefit the sequence generation task used in our study. The task requires the reservoir to produce reproducible, class-specific internal trajectories that unfold over time. Local saturation and fragmentation, as induced by Hopfield reciprocity, limit the reservoir's ability to flexibly traverse such trajectories, and thus lead to a decrease in computational performance.

\subsubsection*{Dale Homogeneity}

\NI Dale's principle, a key feature of biological neural networks, stipulates that each neuron maintains a fixed output polarity—either excitatory or inhibitory—across all its targets. In our model, this principle is implemented by the homogeneity parameter \( h \in [0,1] \), which controls the consistency of output signs within each column of the weight matrix. At \( h = 0 \), neurons send mixed excitatory and inhibitory outputs; at \( h = 1 \), every neuron acts strictly as a monopolar sender.

\NI From a theoretical perspective, one might expect that Dale homogeneity could reduce destructive interference between opposing inputs and introduce a more directional and interpretable signal flow through the reservoir. In particular, we anticipated that increasing \( h \) would attenuate high-frequency or erratic fluctuations by enforcing more coherent influence patterns among neurons.

\NI Our simulations confirm this expectation. As the homogeneity parameter \( h \) increases, we observe a robust and monotonic decrease in the fluctuation measure \( F \), especially in balanced systems (\( b = 0 \)) where chaotic behavior typically dominates. Notably, this stabilization is not accompanied by a major change in the nonlinearity parameter \( N \), indicating that the network remains in a moderately nonlinear regime rather than shifting toward saturation or linearity.

\NI This fluctuation reduction leads to a clear and strong improvement in computational performance. Accuracy in the sequence generation task rises significantly with increasing \( h \), implying that the network becomes more reliable at generating reproducible internal trajectories. Even in slightly unbalanced regimes (\( b > 0 \)), Dale homogeneity continues to support a modest performance gain.

\NI Taken together, these results suggest that Dale's principle promotes more stable and functional dynamics in RNNs—not by altering the degree of nonlinearity, but by reducing internal conflict and noise through output consistency.

\subsubsection*{Modularity (Block Size \( S = 1 \))}

\NI In the case \( S = 1 \), the modularity parameter \( m \) does not produce distinct structural blocks in the weight matrix. Instead, increasing \( m \) gradually transforms the underlying weight distribution from a single Gaussian to a mixture of two Gaussians—one narrower (weaker weights) and one broader (stronger weights)—with the total standard deviation kept constant. This results in a heterogeneous distribution of connection strengths, but without any spatial clustering.

\NI Despite the absence of actual modular structure, this modification already has measurable effects on the system's behavior. As \( m \) increases, we observe a slight decrease in the fluctuation measure \( F \), a moderate reduction in the nonlinearity parameter \( N \), and a clear rise in task accuracy \( A \).

\NI We interpret this as a statistical smoothing effect: by blending strong and weak connections, the network avoids excessive excitation in any particular subset of neurons. This promotes a more moderate operating regime in the reservoir, in which the dynamics remain responsive but do not drive units too deeply into saturation. Consequently, the network operates in a regime that is both less nonlinear and more stable, improving the linear readout's ability to extract task-relevant information.

\subsubsection*{Modularity (Block Size \( S = 10 \))}

\NI In biological neural networks, modular organization is a well-established principle observed across spatial scales—from cortical microcircuits to large brain regions. Modules, typically defined as groups of neurons with strong internal connectivity and weaker coupling to the rest of the network, are thought to support functional specialization while preserving global integration. From this perspective, introducing modularity into artificial RNNs could plausibly enhance their computational capacity by enabling localized processing and buffering against global instabilities.

\NI In our model, modularity is implemented via the control parameter \( m \in [0,1] \), which adjusts the variance of intra- and inter-module connection strengths while keeping the overall standard deviation constant. When the block size is set to \( S = 10 \), the resulting matrix exhibits clearly defined patches of stronger weights embedded in a weaker background. These high-variance patches may lie on or off the diagonal. Diagonal modules imply recurrently coupled local groups, whereas off-diagonal modules encode directional connections from one group of neurons to another, without necessarily reciprocal feedback. While the former can stabilize local loops of activity, the latter may implement feedforward-like interactions across functionally distinct subnetworks. Though less intuitively interpretable, these asymmetric modules still contribute to shaping constrained, layered dynamics within the recurrent architecture.

\NI Prior to numerical investigation, we expected that such modular organization could lead to partial functional segregation, dampening chaotic fluctuations and possibly enabling more reproducible activity patterns within modules. Indeed, these expectations are strongly supported by the simulations. As \( m \) increases, we observe a pronounced reduction in both fluctuation \( F \) and nonlinearity \( N \), indicating that the reservoir dynamics become more stable and less saturated. This effect is especially clear in balanced networks (\( b = 0 \)), which otherwise tend to exhibit chaotic behavior. The drop in \( F \) suggests that spontaneous overactivation is strongly suppressed, while the lower \( N \) points to a shift away from digital-like saturation toward a more linear or intermediate regime.

\NI Most strikingly, the accuracy \( A \) in the sequence generation task increases sharply and reaches near-maximal levels across a wide range of \( m \). This indicates that the modular architecture allows the reservoir to traverse high-dimensional yet structured internal trajectories that can be reliably decoded by the linear readout. The improvement in computational performance surpasses that achieved with Hopfield reciprocity or Dale homogeneity.

\NI We interpret these findings as evidence that modularity imposes a beneficial form of compartmentalization: the strong intra-module coupling stabilizes local activity patterns, while the weak inter-module coupling prevents runaway global synchrony or chaos. Together with the presence of asymmetric inter-module pathways, this creates a balance between dynamical richness and control, facilitating the emergence of reproducible, task-relevant trajectories.

\subsection{Outlook}

\NI The present study has focused exclusively on one specific computational task: the generation of short output sequences from class-specific input stimuli. While this task is well-suited for evaluating the internal stability and reproducibility of reservoir trajectories, it represents only one of many possible functional challenges an RNN might face. Future studies will systematically explore how the three structural regularities—Hopfield reciprocity, Dale homogeneity, and modularity—affect other task types, including classification, prediction, temporal integration, and generative modeling. These tasks may impose different demands on the reservoir, potentially favoring entirely different dynamical regimes.

\NI Several further limitations of the current setup should be noted. Our networks use a fixed, pointwise $\tanh$ activation function throughout, with no mechanism for adapting nonlinear response properties during training or evolution. Moreover, the recurrent connections are entirely fixed, without plasticity or learning mechanisms. While this simplification allows us to isolate the impact of architectural regularities, it remains unclear how the observed effects extend to systems with ongoing synaptic adaptation. Likewise, our networks are relatively small and shallow, lacking hierarchical depth or multi-scale processing pathways. It is an open question whether similar regularities exert comparable influences in large-scale architectures with layered structure, where modularity and reciprocity might interact differently with gradient propagation and representational abstraction.

\NI A further important direction concerns the interaction between structural regularities. In this study, we varied each regularity parameter in isolation while keeping the others fixed. However, real biological systems typically exhibit several regularities at once. It remains an open question whether combinations of regularities act synergistically or interfere with each other. For instance, it is conceivable that modularity and Dale homogeneity together enhance performance more than either alone—or that reciprocity counteracts the benefits of modular organization.

\NI A separate line of inquiry could explore the relevance of the observed effects for biological computation. Since the structural regularities studied here are motivated by neurobiological observations, it is worthwhile to compare their dynamical implications to actual brain circuits. This could involve applying the same dynamical and performance metrics to empirically derived connectomes—such as those of \textit{C. elegans}, \textit{Drosophila}, or the zebrafish larva—and seeing how they perform on comparable tasks.

\NI Beyond empirical validation, the theoretical understanding of how structural regularities shape network dynamics remains incomplete. For instance, it is still unclear why modularity so reliably suppresses fluctuation and nonlinearity, or why Dale homogeneity improves accuracy without driving the system into linearity. While it may be tempting to invoke classical tools such as the spectral radius or eigenvalue spectra of the weight matrix, such linear measures often fail to capture the complex behavior of nonlinear, recurrent systems. In our view, a more fruitful approach would be to characterize

% *******************************************************
\section{Additional Information}
% *******************************************************

\subsection{Author contributions}

CM conceived the study, implemented the methods, evaluated the data, and wrote the paper. PK conceived the study, discussed the results, acquired funding and wrote the paper. AS discussed the results and acquired funding. AM discussed the results and provided resources.

\subsection{Funding}
This work was funded by the Deutsche Forschungsgemeinschaft (DFG, German Research Foundation): KR\,5148/3-1 (project number 510395418), KR\,5148/5-1 (project number 542747151), and GRK\,2839 (project number 468527017) to PK, and grant SCHI\,1482/3-1 (project number 451810794) to AS.

\subsection{Competing interests statement}
The authors declare no competing interests.

\subsection{Data availability statement} 
The complete data and analysis programs will be made available upon reasonable request.

\subsection{Third party rights}
All material used in the paper are the intellectual property of the authors.

% ***************************************************
% REFERENCES
% ***************************************************

\newpage
%\bibliographystyle{unsrt}
%\bibliography{references}

% ***************************************************
% FIGURES
% ***************************************************

% -------------------------------------------------
\newpage
% -------------------------------------------------

% -------------------------------------------------
\newpage
% -------------------------------------------------

% -------------------------------------------------
\newpage
% -------------------------------------------------

% -------------------------------------------------
\newpage
% -------------------------------------------------

% -------------------------------------------------
\newpage
% -------------------------------------------------

\end{document}